%% file: Optomagnonic_Cooling.tex
\documentclass[reprint,aps,prl,superscriptaddress,nobibnotes]{revtex4-1}
\usepackage{amsmath}
\usepackage{amssymb}
\usepackage{natbib}
\usepackage{graphicx}
\usepackage{xcolor}

\setcitestyle{numbers,square}

\begin{document}

\title{Optical cooling of magnons} 

\author{Sanchar Sharma} 
\affiliation{Kavli Institute of NanoScience, Delft University of Technology, 2628 CJ Delft, The Netherlands}  

\author{Yaroslav M. Blanter} 
\affiliation{Kavli Institute of NanoScience, Delft University of Technology, 2628 CJ Delft, The Netherlands}  

\author{Gerrit E. W. Bauer}  
\affiliation{Institute for Materials Research \& WPI-AIMR \& CSRN, Tohoku University, Sendai 980-8577, Japan} 
\affiliation{Kavli Institute of NanoScience, Delft University of Technology, 2628 CJ Delft, The Netherlands}  

\date{\today}

\begin{abstract}
Inelastic scattering of light by spin waves generates an energy flow between the light and magnetization fields, a process that can be enhanced and controlled by concentrating the light in magneto-optical resonators. Here, we model the cooling of a sphere made of a magnetic insulator, such as yttrium iron garnet (YIG), using a monochromatic laser source. When the magnon lifetimes are much larger than the optical ones, we can treat the latter as a Markovian bath for magnons. The steady-state magnons are canonically distributed with a temperature that is controlled by the light intensity. We predict that such a cooling process can significantly reduce the temperature of the magnetic order within current technology.
\end{abstract}

\maketitle  

A great achievement of modern physics is the Doppler cooling of trapped atoms by optical lasers \cite{HanschSchalow,WilliamNobel} down to temperatures of micro-Kelvin \cite{Lett88}. Subsequently, even macroscopic mechanical objects, such as membranes and cantilevers, have been cooled to their quantum mechanical ground state \cite{Florian07,Wilson07,Connell10,Chan11,Aspelmeyer14} by blue shifting the stimulated emission using an optical cavity \cite{Florian07,Wilson07}. `Cavity optomechanics' is a vibrant field that achieved successful Heisenberg uncertainty-limited mechanical measurements, the generation of entangled light-mechanical states, and ultra-sensitive gravitational wave detection \cite{Aspelmeyer14}. An optical cryocooler based on solid state samples \cite{Edwards99} can be superior due to its compactness and lack of moving components \cite{SBELaserCoolingSolids}. Optical cooling has been demonstrated for glass \cite{Epstein95,Edwards99} and envisioned for semiconductors \cite{SBESemiconductors,SBELaserCoolingSolids,Usami12}.  

An analogous cooling of a magnet would generate interesting opportunities. Magnetization couples to microwaves \cite{SoykalPRL10,Zhang14,TabuchiHybrid14,Zhang16,Bai15}, electric currents \cite{Kajiwara10,Chumak15,Bai15}, mechanical motion \cite{Kittel58,Akash15,Kikkawa16,Zhang16}, and, indeed, light \cite{Borovik82}. Spin waves are the elementary excitations of the ferromagnetic order, which are quantized as bosonic magnons. Similar to phonons, magnons may be considered non-interacting up to relatively high temperatures and are Planck-distributed at thermal equilibrium. However, there are important differences as well: Magnons have mass and chirality \cite{DamEshSlab,StPrMagnons}, both of which are tunable by an external static magnetic field. Their long wavelength dispersion in thin films is highly anisotropic, with minima in certain directions that can collect the Bose-Einstein condensate of magnons \cite{DemokritovBEC,RezendeBEC,BECBook}. Magnons can be used as quantum transducers between microwaves and optical light \cite{Hisatomi16} or between superconducting and flying qubits \cite{TabuchiQubit15}.  

Motivated by the potential of a ferromagnet as a versatile quantum interface at low temperatures, we discuss here the potential of optical cooling of magnons. The magnon-photon interaction gives rise to inelastic Brillouin light scattering (BLS) \cite{StPrMO}, which is a well established tool to study magnon dispersion and dynamics \cite{Borovik82,Sebastian15Rev,Klingler16}. Recently, several groups carried out BLS experiments on spheres made of ferrimagnetic insulator yttrium iron garnet (YIG) \cite{James15,Osada16,ZhangWGM16,JamesWGM,Osada17_Exp,JamesLowL}, which has a very high magnetic quality factor $\left( \sim \mathrm{10}^5\right) $ \cite{Cherepanov_YIG,Serga10,WuHoff} and supports ferromagnetic-like magnons with long coherence times $\left( \sim \mathrm{\mu}\text{s}\right) $ \cite{ZhangMemory15,TabuchiQubit15,Tabuchi16}. YIG spheres are commercially available for microwave applications, but are also good infrared light cavities due to their large refractive index and high optical quality \cite{WettlingData76,LacklisonBiYIG}, making them good optomagnonic resonators  \cite{James15,Osada16,ZhangWGM16,JamesWGM,Tianyu16,Silvia16,WGMOptoMag,Osada17_Exp,Osada17_Th}. Via proximity optical fibers or prisms, external laser light can efficiently excite `whispering gallery modes' (WGMs), i.e. the optical modes circulating in extremal orbits of dielectric spheroids \cite{Oraevsky02,Foreman15}.  

The BLS experiments on YIG spheres discovered a large asymmetry in the red- (Stokes) and blue-shifted (anti-Stokes) sidebands \cite{JamesWGM,ZhangWGM16,Osada16,Osada17_Exp} due to selective resonant enhancement of the scattering cross section \cite{JamesWGM,WGMOptoMag,Osada17_Th}. The asymmetry can be controlled by the polarization and wave vector of the light. When more photons are scattered into the blue than the red-shifted sidebands, light effectively extracts energy from the magnons. Optomagnonic scattering is enhanced for a `triple resonance condition' \cite{Dobrindt10,Amir11,Mari12,Bahl12,JamesWGM} by tuning both the input and the scattered photon frequency to the optical resonances of the cavity. In contrast, optomechanical cooling \cite{Wilson07,Florian07,Aspelmeyer14} requires detuning the input laser from a cavity resonance with correspondingly reduced scattering and cooling rate. In this manuscript, we predict that modern technology and materials can significantly reduce the temperature of the magnetic order, showing the potential to manipulate magnons using light.

\begin{figure}[tbp]
	\[ \includegraphics[width=.47\textwidth,keepaspectratio]{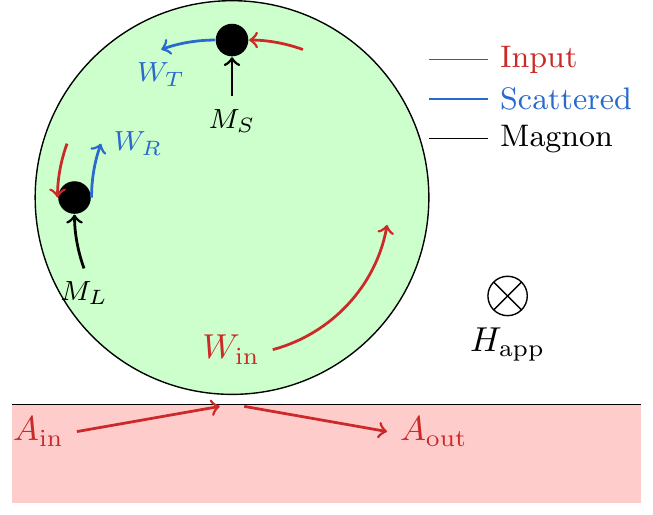} \]
	\caption{Optomagnonic cooling setup: A ferromagnetic sphere in contact with an optical waveguide. A magnetic field $H_{\text{app}}$ (into the paper) is applied to saturate the magnetization. Input light with amplitude $A_{\mathrm{in}}$ is evanescently coupled to a WGM $W_{\mathrm{in}}$. We focus on anti-Stokes scattering by two types of magnons that are characterized by their angular momentum \cite{WGMOptoMag}. A small angular momentum magnon $M_S$ maintains the direction of WGMs, converting $W_{\mathrm{in}}$ to $W_{T}$. $W_{\mathrm{in}}$ can be reflected into $W_{R}$ by absorbing a large angular momentum magnon $M_L$. Theoretically, both the cases can be treated in the same formalism. } \label{Fig:Setup}
\end{figure}

We derive below rate equations for photons and magnons to estimate the steady-state magnon number that can be reached as a function of material and device parameters. We consider a spherical magnetic insulator with high index of refraction that is transparent at the input light frequency (Fig.\thinspace \ref{Fig:Setup}) and magnetization perpendicular to the WGM orbits that are excited by proximity coupling to an external laser. We single out two groups of magnon modes that couple preferentially to the WGMs \cite{WGMOptoMag}. The small angular momentum (including the Kittel) magnons, $ M_S $ in Fig.\,\ref{Fig:Setup}, and large angular momentum magnons, the chiral Damon-Eshbach (DE) modes $M_L$. The theory presented below is valid for both types of magnons.

\begin{figure}[tbp]
	\[ \includegraphics[width=.45\textwidth, keepaspectratio]{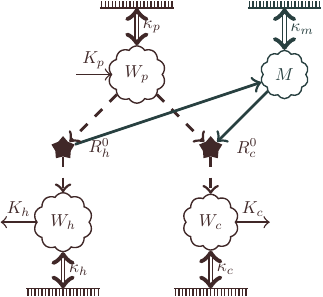} \]
	\caption{Light-induced cooling of a magnon, $M$. A proximity fiber or prism is coupled to the WGMs $W_x$ with a coupling constant $K_x$, exciting $W_p$ while collecting the scattered $W_c$ and $W_h$. The photons are inelastically scattered by the magnon $W_{p}+M\rightarrow W_{c}$ and $ W_{p}\rightarrow W_{h}+M$ at single particle rates $R_{c}^{0}$ and $R_{h}^{0}$ respectively, derived in the text. All modes are coupled to their respective thermal baths by leakage rates $\kappa_{x}$. When $\kappa_{c}$ is much larger than the corresponding scattering rate, the bath associated with $W_{c}$ can become an efficient channel for dissipation of the magnons in $M$. } \label{Fig:ScatBasics}
\end{figure}

We can understand the basic physics by the minimal model sketched in Fig.\,\ref{Fig:ScatBasics}. We focus on a single incident WGM $W_p$ with index $p$ and frequency $\omega_p$. It is occupied by \cite{Aspelmeyer14}  
\begin{equation}
	 n_p = \frac{4K_p}{(\kappa_p + K_p)^2}\frac{P_{\mathrm{in}}}{\hbar \omega_p} \label{Def:np} 
\end{equation} 
photons, with $\kappa_p$ being the intrinsic linewidth, $K_p$ the leakage rate into the proximity coupler, and $P_{\mathrm{in}}$ the input light power. An optically active magnon $M$ [with either small or large angular momentum] is annihilated $W_p + M\rightarrow W_c$ or created $W_p\rightarrow W_h + M$ by BLS, where $W_c$ and $W_h$ are blue and red-shifted sideband WGMs, respectively.  

We first derive a simple semi-classical rate equation for the non-equilibrium steady-state magnon number, $n_m^{(sc)}$ [the superscript distinguishes the estimate from $n_m$ as more rigorously derived below]. The thermal bath absorbs and injects magnons at rates $\kappa_mn_m^{(sc)} \left( n_{\mathrm{th}} + 1 \right)$ and $ \kappa_mn_{\mathrm{th}} \left( n_m^{(sc)} + 1 \right)$ respectively, where $\kappa_m$ is the inverse magnon lifetime,  
\begin{equation}
	 n_{\mathrm{th}} = \left[ \exp \left( \frac{\hbar \omega_m}{k_BT}\right) - 1 \right]^{- 1} \label{Def:nmth} 
\end{equation} 
its equilibrium thermal occupation, $\omega_m$ the magnon frequency and $T $ the ambient temperature. The optical cooling rate is $ R_c^0 n_p n_m^{(sc)}$, where $R_c^0$ is the anti-Stokes scattering rate of one $W_p$-photon by one $M$-magnon and we assumed that there are no photons in $W_c$. The latter is justified because of small optomagnonic couplings compared to WGM dissipation rates, $\sim 2\pi \times 0 . 1 - 1\,$GHz \cite{JamesWGM,Osada16,ZhangWGM16} while $R_c^0 n_p n_m^{(sc)}$ is at most $\sim \kappa_m \sim 2\pi \times 1$ \thinspace MHz. In the absence of dissipation, Fermi's golden rule gives $R_c^0 = 2\pi |g_c|^2\delta (\omega_p + \omega_m - \omega_c)$, where $\hbar g_c$ is the optomagnonic coupling and $ \{\omega_p,\omega_c,\omega_m\}$ are the frequencies of $ \{W_p,W_c,M\},$ respectively. When $W_c$ has a finite lifetime, the $ \delta $-function is broadened into a Lorentzian, giving
\begin{equation}
	 R_c^0 = \frac{|g_c|^2 (\kappa_c+K_c)}{\left( \omega_p + \omega_m - \omega_c\right)^2 + (\kappa_c + K_c)^2/4}, \label{Def:Rc} 
\end{equation} 
where $\kappa_c$ is its intrinsic linewidth, and $K_c$ is its leakage rate into the proximity coupler. Similarly, the optical heating rate is $ R_h^0n_p \left(n_m^{(sc)} + 1\right),$ where $R_h^0$ is given by Eq.\,(\ref{Def:Rc}) with $ g_c,\omega_c,\kappa_c\rightarrow g_h,\omega_h,\kappa_h$ and $ \omega_m\rightarrow - \omega_m$. In deriving $R_{c,h}^0$, we ignore the magnon linewidth since $\kappa_m\ll \kappa_c,\kappa_h$ \cite{JamesWGM,WGMOptoMag}. In the steady state the cooling and heating rates are equal, leading to the estimate
\begin{equation}
	 n_m^{(sc)} = \frac{\kappa_mn_{\mathrm{th}} + R_h^0n_p}{\kappa_m + \left(R_c^0 - R_h^0\right) n_p}. \label{Class:nm} 
\end{equation} 
This agrees with the result from the more precise theory discussed below, thus capturing the essential processes correctly (a posteriori). However, the rate equation cannot access noise properties beyond the magnon number that are important for thermodynamic applications. Further, it does not differentiate between a coherent precession of the magnetization and the thermal magnon cloud with the same number of magnons.

In order to model the cooling process more rigorously, we proceed from a model Hamiltonian for a system with three photon and one magnon mode. In the Hamiltonian $\hat{H}_S = \hat{H}_0 + \hat{H}_{\mathrm{om}}$ \cite{WGMOptoMag} 
\begin{equation}
	 \hat{H}_0 = \hbar \omega_p\hat{a}_p^{\dagger}\hat{a}_p + \hbar \omega_c\hat{a}_c^{\dagger}\hat{a}_c + \hbar \omega_h\hat{a}_h^{\dagger} \hat{a}_h + \hbar \omega_m\hat{m}^{\dagger}\hat{m}, 
\end{equation} 
and $\hat{a}_x$ and $\hat{m}$ are the annihilation operators for photon $W_x$ with $x\in \left\{p,c,h\right\} $ and magnon $M$. The optomagnonic coupling in the rotating wave approximation reads \cite{WGMOptoMag} 
\begin{equation}
	 \hat{H}_{\mathrm{om}} = \hbar g_c\hat{a}_p\hat{a}_c^{\dagger}\hat{m} + \hbar g_h\hat{a}_p\hat{a}_h^{\dagger}\hat{m}^{\dagger} + \mathrm{h . c .} , 
\end{equation} 
where $g_c$ and $g_h$ are the scattering amplitudes and $\mathrm{h . c .}$ is the Hermitian conjugate.  

In the rotating frame of the \textquotedblleft envelope\textquotedblright\ operators $\hat{W}_x(t)\overset{\triangle}{=}\hat{a}_x(t)e^{i\omega_xt}$ and $\hat{M}(t)\overset{\triangle}{=}\hat{m}(t)e^{i\omega_mt}$ the (Heisenberg) equation of motion for $\hat{M}$ becomes \cite{GardinerOrig,Florian07}  
\begin{equation}
	 \dot{\hat{M}} =  - ig_h\hat{W}_p\hat{W}_h^{\dagger}e^{i\delta_ht} - ig_c^{\ast}\hat{W}_p^{\dagger}\hat{W}_ce^{- i\delta_ct} - \frac{\kappa_m}{2}\hat{M} - \sqrt{\kappa_m}\hat{b}_m, \label{Magnon:FirstEOM} 
\end{equation} 
where $\delta_h = \omega_h + \omega_m - \omega_p$ and $\delta_c = \omega_c - \omega_m - \omega_p$ are the detunings from the scattering resonances. $\hat{b}_m(t)$ is the stochastic magnetic field generated by the interaction of $M$ with phonons \cite{Vittoria10} and/or other magnons \cite{Schlomann58}, whose precise form depends on the microscopic details of the interaction \cite{RivasHuelga}. 

We assume that the correlators of $\hat{b}_m$ obey the fluctuation-dissipation theorem for thermal equilibrium \cite{BrownOrig,LL-FDT}. When $\kappa_m \ll k_BT/\hbar$, which is satisfied for $\kappa_m \sim 2\pi \times 1$ \thinspace MHz \cite{JamesWGM,Osada16,ZhangWGM16} and $T \gg 50 \mu$K, the (narrow band filtered) noise is effectively white and generates a canonical Gibbs distribution of the magnons in steady state \cite{GardinerOrig}. Their statistics are $\left\langle \hat{b}_m(t)\right\rangle = 0$, $\left\langle \hat{b}_m^{\dagger}(t^{\prime}) \hat{b}_m(t)\right\rangle = n_{\mathrm{th}}\delta (t - t^{\prime})$ and $ \left\langle \hat{b}_m(t^{\prime})\hat{b}_m^{\dagger}(t)\right\rangle = (n_{\mathrm{th}} + 1)\delta (t - t^{\prime})$, where $n_{\mathrm{th}}$ is defined in Eq.\thinspace (\ref{Def:nmth}). 

For weak scattering relative to the input power, we can ignore any back-action on $\hat{W}_p$ such that its dynamics is governed only by the proximity coupling. When $W_p$ is in a coherent state, $\left\langle \hat{W}_p(t)\right\rangle = \sqrt{n_p}$ and $\left\langle \hat{W}_p^{\dagger}(t^{\prime})\hat{W}_p(t)\right\rangle = n_p$, where $n_p$ is given by Eq.\thinspace (\ref{Def:np}).  

The photons in $W_c$ are generated by $\hat{H}_{\mathrm{om}}$ and dissipated into their thermal bath, with Heisenberg equation of motion \cite{GardinerOrig,Florian07}  
\begin{equation}
	 \frac{d\hat{W}_c}{dt} =  - ig_c\hat{W}_p\hat{M}e^{i\delta_ct} - \frac{\kappa_c+K_c}{2}\hat{W}_c - \sqrt{\kappa_c}\hat{b}_c - \sqrt{K_c}\hat{A}_c , \label{Wc:FirstEOM} 
\end{equation} 
where $\hat{b}_c$ and $\hat{A}_c$ are noise operators. The physical origins of $\hat{b}_c$ and finite lifetime $\kappa_c^{- 1}$ are scattering by impurities, surface roughness, and lattice vibrations. $K_c$ is the leakage rate of $W_c$ into the proximity coupler and $\hat{A}_c$ is the vacuum noise from the latter into $W_c$. The noise sources are white for sufficiently small $\kappa_c$. $\left\langle \hat{X}_c(t)\right\rangle = 0$, $\left\langle \hat{X}_c^{\dagger}(t^{\prime}) \hat{X}_c(t)\right\rangle = 0$ and $\left\langle \hat{X}_c(t^{\prime}) \hat{X}_c^{\dagger}(t)\right\rangle = \delta (t - t^{\prime})$ where $X \in \{\hat{b}_c,\hat{A}_c\}$, because the thermal occupation of photons at infrared and visible frequencies is negligibly small at room temperature $e^{- \hbar \omega_c/(k_BT)} \approx 0$.  

The solution to Eq.\thinspace (\ref{Wc:FirstEOM}) is $\hat{W}_c(t) = \hat{W}_{c,\mathrm{th}}(t) + \hat{W}_{c,\mathrm{om}}(t)$. The thermal contribution is,  
\begin{equation}
	 \hat{W}_{c,\mathrm{th}} = \int_{0}^t e^{- (\kappa_c+K_c) (t-\tau)/2} \left[-\sqrt{\kappa_c} \hat{b}_c(\tau) - \sqrt{K_c}\hat{A}_c(\tau) \right] d\tau \label{Wc:Th} 
\end{equation} 
where the origin of time is arbitrary. For $t,t' \rightarrow \infty$, we get the equilibrium statistics $\left\langle \hat{W}_{c,\mathrm{th}}^{\dagger}(t^{\prime})\hat{W}_{c, \mathrm{th}}(t)\right\rangle = 0$ and  
\begin{equation}
	 \left\langle \hat{W}_{c,\mathrm{th}}(t^{\prime})\hat{W}_{c,\mathrm{th}}^{\dagger}(t)\right\rangle = \exp\left[-\frac{(\kappa_c + K_c)|t - t^{\prime}|}{2}\right], \label{Wc:Th:Corr} 
\end{equation} 
independent of the optomagnonic coupling. $\hat{W}_{c,\text{\textrm{om}}}$ can be simplified by the adiabaticity of the magnetization dynamics that follows from $\kappa_m\ll \kappa_c$. When $\hat{M}$ is treated as a slowly varying constant  
\begin{equation}
	 \hat{W}_{c,\mathrm{om}}(t) \approx  - ig_c\hat{M}(t)\int_{0}^te^{-(\kappa_c+K_c)(t-\tau)/2} \hat{W}_p(\tau) e^{i\delta_c\tau} d\tau . \label{Wc:OM} 
\end{equation} 
$\hat{W}_h(t)$ is obtained by the substitution $c\rightarrow h$ and $\hat{M}\rightarrow \hat{M}^{\dagger}$ in Eqs. (\ref{Wc:Th}-\ref{Wc:OM}).

We can now rewrite Eq.\,(\ref{Magnon:FirstEOM}) as  
\begin{equation}
	 \frac{d\hat{M}}{dt} =  - \left( \frac{\kappa_m}{2}\hat{M} + \sqrt{\kappa_m} \hat{b}_m\right) + \hat{O}_c + \hat{O}_h . \label{Magnon:Env} 
\end{equation} 
with cooling and heating operators that reflect the light scattering processes in Fig.\,\ref{Fig:ScatBasics}:  
\begin{align}
	\hat{O}_{c} &= \hat{\mathcal{N}}_{c} + i\hat{\Sigma}_{c}\hat{M}, \\
	\hat{O}_{h} &= -\hat{\mathcal{N}}_{h}^{\dagger} + i\hat{\Sigma}_{h}^{\dagger} \hat{M}.
\end{align}

Focusing on the cooling process, we distinguish the self-energy,  
\begin{equation}
	 \hat{\Sigma}_c = i|g_c|^2 \int_{0}^te^{(i\delta_c + (\kappa_c + K_c)/2)(\tau - t)}\hat{W}_p^{\dagger}(t)\hat{W}_p(\tau )d\tau , \label{Def:SEc} 
\end{equation} 
from the noise operator,  
\begin{equation}
	 \hat{\mathcal{N}}_c(t) =  - ig_c^{\ast}\hat{W}_p^{\dagger}(t)\hat{W}_{c, \mathrm{th}}(t)e^{- i\delta_ct} . \label{Def:Nc} 
\end{equation} 
In the weak-coupling regime we may adopt a mean-field approximation by replacing $\hat{\Sigma}_c$ by its average,  
\begin{equation}
	 \left\langle \hat{\Sigma}_c\right\rangle = -\bar{\omega}_c + i\frac{\bar{\kappa}_c}{2}\overset{\triangle}{=} \frac{|g_c|^2n_p}{\delta_c - i(\kappa_c+K_c)/2}, \label{InMagDefs} 
\end{equation} 
where $\bar{\omega}_c$ is the (reactive) shift of the magnon resonance and $ \bar{\kappa}_c$ the optical contribution to the magnon linewidth.

The noise $\hat{\mathcal{N}}_c$ can be interpreted as the vacuum fluctuations of $W_c$ entering the magnon subsystem via the optomagnonic interaction. $\hat{\mathcal{N}}_c$ has a very short correlation time $\sim (\kappa_c+K_c)^{- 1}$ [see Eq.\thinspace (\ref{Wc:Th:Corr})] compared to magnon dynamics $\sim \kappa_m^{- 1}$, and thus can be treated as a white noise source with $\left\langle \hat{\mathcal{N}}_c(t)\right\rangle = 0$, $ \left\langle \hat{\mathcal{N}}_c^{\dagger}(t)\hat{\mathcal{N}}_c(t^{\prime})\right\rangle = 0$ , and $\left\langle \hat{\mathcal{N}}_c(t^{\prime})\hat{\mathcal{N}}_c^{\dagger}(t)\right\rangle \approx V_c \delta (t - t^{\prime})$. By integrating over time and using the correlation functions of $\hat{W}_p$ and $\hat{W}_{c,\mathrm{th}}$  
\begin{equation}
	 V_c = \frac{4|g_c|^2n_p (\kappa_c+K_c)}{4\delta_c^2 + (\kappa_c+K_c)^2} = \bar{\kappa}_c , \label{Nc:White} 
\end{equation} 
defined in Eq.\,(\ref{InMagDefs}). $\bar{\kappa}_c/\kappa_m$ at resonance $\delta_c=0$ is the cooperativity between the magnons and $W_c$-photons due to the coupling mediated by $W_p$-photons.

Analogous results hold for $\hat{O}_h$, with substitutions $c\rightarrow h$ in Eqs.\,(\ref{Def:SEc})-(\ref{Nc:White}). We arrive at  
\begin{equation}
	 \frac{d\hat{M}}{dt} \approx  - i(\bar{\omega}_c + \bar{\omega}_h)\hat{M} - \frac{\kappa_{\mathrm{tot}}}{2}\hat{M} - \sqrt{\kappa_{\mathrm{tot}}}\hat{b}_{\mathrm{tot}}, \label{Mag:WN} 
\end{equation} 
where $\kappa_{\mathrm{tot}} = \kappa_m + \bar{\kappa}_c - \bar{\kappa}_h$ and $\sqrt{\kappa_{\mathrm{tot}}}\hat{b}_{\mathrm{tot}} = \sqrt{\kappa_m} \hat{b}_m - \hat{\mathcal{N}}_c + \hat{\mathcal{N}}_h^{\dagger}$. The fluctuations of the total noise follow from Eq.\thinspace (\ref{Nc:White})  
\begin{align}
	\left\langle \hat{b}_{\mathrm{tot}}^{\dagger}\left(t^{\prime}\right) \hat{b}_{\mathrm{tot}}\left(t\right) \right\rangle &\approx n_{m}\,\delta\left(t-t^{\prime}\right), \\
	\left\langle \hat{b}_{\mathrm{tot}}\left(t^{\prime}\right) \hat{b}_{\mathrm{tot}}^{\dagger}\left(t\right) \right\rangle &\approx \left(n_{m}+1\right)\,\delta\left(t-t^{\prime}\right),
\end{align}
where  
\begin{equation}
	 n_m = \frac{\kappa_mn_{\mathrm{th}} + \bar{\kappa}_h}{\kappa_m + \bar{\kappa}_c - \bar{\kappa}_h}. \label{Def:nm} 
\end{equation}

Eq.\thinspace (\ref{Mag:WN}) is equivalent to the equation of motion for magnons in equilibrium after the substitutions $\omega_m\rightarrow \omega_m + \bar{\omega}_c + \bar{\omega}_h$, $\kappa_m\rightarrow \kappa_{\mathrm{tot}}$, and $n_{\mathrm{th}}\rightarrow n_m$. It implies that the magnons in the non-equilibrium steady state are still canonically distributed with density matrix  
\begin{equation}
	 \hat{\rho}_{ne} = \exp\left(\frac{-\hbar \omega_m\hat{n}_m}{k_BT_{ne}}\right) \left(\mathrm{Tr}\left[ \exp\left(\frac{-\hbar \omega_m\hat{n}_m}{k_BT_{ne}}\right)\right]\right)^{-1}
\end{equation} 
where the number operator $\hat{n}_m = \hat{m}^{\dagger}\hat{m}$ and the non-equilibrium magnon temperature $T_{ne}$ is implicitly defined by Eq.\,(\ref{Def:nm}) and
\begin{equation}
	 n_m = \left[ \exp\left(\frac{\hbar\omega_m}{k_BT_{ne}} \right) - 1\right]^{-1}. \label{Bose} 
\end{equation} 
 
We get $\left\langle \hat{M}_x \right\rangle = \left\langle \hat{M}_y \right\rangle = 0$, which implies that light scattering does not induce a coherent magnon precession, in contrast to a resonant ac magnetic field. $n_m$ is the average number of magnons that can be larger or smaller than the equilibrium value $n_{\mathrm{th}}$. The result is consistent with $n_m^{(sc)}$ [see Eq.\,(\ref{Class:nm})] because $\bar{\kappa}_{c,h} = R_{c,h}^0 n_p$ as expected from Fermi's golden rule. The canonical distribution implies that the steady-state magnon entropy is maximized for the given number of magnons, $n_m$. 

When $\bar{\kappa}_h - \bar{\kappa}_c>\kappa_m,$ i.e. when heating by the laser overcomes the intrinsic magnon damping, the system becomes unstable, leading to runaway magnon generation and self-oscillations \cite{SuhlOrig,Rezende90,Silvia16}. The instability is regularized by magnon-magnon scattering, not included in our theory. 

Here we focus on the cooling scenario in which $\bar{\kappa}_h\ll \bar{\kappa}_c$ \cite{WGMOptoMag}. Magnon cooling can be monitored by the intensity of the blue-shifted sideband. Using the input-output formalism \cite{GardinerOrig,Clerk10} the scattered light amplitude in the rotating frame is
\begin{equation}
	 \hat{A}_{\mathrm{out}}(t) =  - \sqrt{K_c}\hat{W}_c(t). 
\end{equation} 
It can be converted into the output power by $P_{\mathrm{out}} = \hbar \omega_c\left\langle \hat{A}_{\mathrm{out}}^{\dagger}(t) \hat{A}_{\mathrm{out}}(t) \right\rangle $, which is independent of time in steady state. With impedance matching, $\kappa_{p,c} = K_{p,c} $, and at the triple resonance, $\delta_c = 0$,  
\begin{equation}
	 \frac{P_{\mathrm{out}}}{P_{\mathrm{in}}} = \frac{\left|g_c\right|^2}{\kappa_c\kappa_p} \frac{\kappa_mn_{\mathrm{th}}}{\kappa_m + 2|g_c|^2n_p/\kappa_c}\propto \frac{1}{1 + P_{\mathrm{in}}/P_s}, 
\end{equation} 
defining the saturation power  
\begin{equation}
	 P_s\overset{\triangle}{=} \frac{\hbar \omega_p\kappa_p\kappa_c\kappa_m}{2|g_c|^2} . \label{Def:Ps} 
\end{equation} 
To leading order $P_{\mathrm{out}}\propto P_{\mathrm{in}}$ \cite{ZhangWGM16,WGMOptoMag}, but saturates when the magnon number becomes small, which is an experimental evidence for magnon cooling. $P_s$ is the input power that halves the number of magnons.  

For a YIG sphere with parameters $\omega_c \approx \omega_p = 2\pi \times 300\,$THz (free space wavelength $1\mu$m), an optical Q-factor $\omega_p/(2\kappa_p) = \omega_c/(2\kappa_c) = 10^6$, \cite{ZhangWGM16}, magnon linewidth $\kappa_m = 2\pi \times 1\,$MHz, and optomagnonic coupling $g_c = 2\pi \times 10\,$Hz \cite{WGMOptoMag}, we get $P_s = 140\,$W. Trying to match this with $P_{\mathrm{in}}$ is not useful since laser-induced lattice heating \cite{SBELaserCoolingSolids} will overwhelm the cooling effect. However, $P_s$ can be significantly reduced by large magnon-photon coupling. Doping YIG with bismuth can increase $g_c$ tenfold  \cite{LacklisonBiYIG}, bringing $P_s$ down to $\sim 1\,$W. The spatial overlap between the magnons and photons \cite{WGMOptoMag} can be engineered in ellipsoidal or nanostructured magnets \cite{Heyroth18} which can increase $g_c$ further by an order of magnitude, giving $P_s \sim 10$mW. For an ambient temperature $ T = 1\,\mathrm{K}$ and magnon frequency $\omega_m = 2\pi \times 10\,$GHz, the thermal magnon number $n_{\mathrm{th}} = 1.62$. For $P_{\mathrm{in}} = \{P_s/20,\,P_s,\,5P_s\}$ the steady-state magnon numbers are $ n_m = \{1 . 55,\,0 . 81,\,0 . 27\}$ and temperatures $ T_{ne} = \{0 . 96,\,0 . 60,\,0 . 31\}\,\mathrm{K}$ respectively. At an optimistic $P_s = 10$mW, the above input power corresponds to $n_p = \{3\times 10^6,\,5\times10^7,\,3\times 10^8\}$ intra-cavity photons respectively. Cooling is experimentally observable for relatively small powers $P_{\mathrm{in}}<P_s/20$, which should be achievable by optimising the magnon-photon coupling.

In summary, we estimate the cooling power due to BLS of light by magnons in an optomagnonic cavity. Due to the large mismatch of optical and magnonic time scales, the photon degree of freedom can be eliminated by renormalizing the magnon frequency and damping, cf. Eq.\thinspace (\ref{Mag:WN}), and a light-controlled effective temperature Eq.\thinspace (\ref{Def:nm}). Current technology and materials are close to achieving significant cooling of magnons, envisioning the possibility of light-controlled magnon manipulation.  

\begin{acknowledgments}
We thank J. Haigh, M. Elyasi, K. Satoh, K. Usami, and A. Gloppe for helpful inputs and discussions. This work is financially supported by the Nederlandse Organisatie voor Wetenschappelijk Onderzoek (NWO) as well as JSPS KAKENHI Grant Nos. 26103006.
\end{acknowledgments}

\input{Optomagnonic_Cooling.bbl}

\end{document}

%% file: Optomagnonic_Cooling.bbl
%